\documentclass[iop]{emulateapj}
\RequirePackage{epstopdf}
\usepackage{multirow}
\RequirePackage{graphicx}
\usepackage{lineno}

\newcommand{\kepler}{{\it Kepler }}
\newcommand{\teff}{$T_{\mathrm{eff}}$}

\newcommand{\um}{$\mu$m}

\slugcomment{Revised for ApJ on \today}

\shorttitle{Metallicities of \kepler M dwarfs}
\shortauthors{Mann et al.}

\begin{document}

\title{Testing the Metal of Late-Type Kepler Planet Hosts with Iron-Clad Methods}

\author{Andrew W. Mann\altaffilmark{1}, Eric Gaidos\altaffilmark{2}, Adam Kraus\altaffilmark{3}, Eric J. Hilton\altaffilmark{1}}
  
\altaffiltext{1}{Institute for Astronomy, University of Hawai'i, 2680 Woodlawn Dr, Honolulu, HI 96822} 
\altaffiltext{2}{Department of Geology \& Geophysics, University of Hawai'i, 1680 East-West Road, Honolulu, HI 96822} 
\altaffiltext{3}{Clay Fellow; Harvard-Smithsonian Center for Astrophysics, 60 Garden St, Cambridge, MA 02138, USA}

\begin{abstract}
It has been shown that F, G, and early K dwarf hosts of Neptune-sized planets are not preferentially metal-rich. However, it is less clear whether the same holds for late K and M dwarf planet hosts. We report metallicities of \kepler targets and candidate transiting planet hosts with effective temperatures below $4500$~K. We use new metallicity calibrations to determine [Fe/H] from visible and near-infrared spectra. We find that the metallicity distribution of late K and M dwarfs monitored by \kepler is consistent with that of the solar neighborhood. Further, we show that hosts of Earth- to Neptune-sized planets have metallicities consistent with those lacking detected planets and rule out a previously claimed 0.2~dex offset between the two distributions at $6\sigma$ confidence. We also demonstrate that the metallicities of late K and M dwarfs hosting multiple detected planets are consistent with those lacking detected planets. Our results indicate that multiple terrestrial and Neptune-sized planets can form around late K and M dwarfs with metallicities as low as 0.25 of the solar value. The presence of Neptune-sized planets orbiting such low-metallicity M dwarfs suggests that accreting planets collect most or all of the solids from the disk and that the potential cores of giant planets can readily form around M dwarfs. The paucity of giant planets around M dwarfs compared to solar-type stars must be due to relatively rapid disk evaporation or a slower rate of core accretion, rather than insufficient solids to form a core.
\end{abstract}



\section{Introduction}\label{sec:intro}
The NASA {\it Kepler} mission \citep{Borucki:2010lr} has discovered more than 2000 exoplanet candidates \citep[also called \kepler Objects of Interest or KOIs, ][]{Batalha:2013lr}, enabling the study of exoplanet statistics based on large data sets. Among other science results, \kepler data has been used to estimate planet occurrence \citep[e.g.,][]{Howard:2012yq, Fressin:2013qy}, constrain the distribution of planet densities \citep{Gaidos:2012lr, Wolfgang:2012uq}, study the architecture of multi-planet systems \citep{Fabrycky:2012lr}, and search for correlations (or non-correlations) between the radius of planets and the metallicity of their host stars for F, G, and early K dwarfs \citep{2012Natur.486..375B}. 

It is well established that the presence of Jovian planets is correlated with the metallicity of the host star \citep{1997MNRAS.285..403G, Santos:2001rt, Fischer:2005yq}. This is generally interpreted as supporting the core accretion mechanism for giant planet formation; metal-rich stars are assumed to have had metal-rich disks in which the higher density of solids allowed faster core accretion and the formation of giant planets before the disks dissipated. \citet{2012Natur.486..375B} showed that Earth- to Neptune-sized planets are present around FGK dwarfs spanning a range of metallicities ($-0.6<$\,[Fe/H]\,$<+0.5$). However, \citet{2012Natur.486..375B} did not measure the metallicity of a control sample of \kepler field stars. If \kepler is biased towards more metal-poor stars (compared to the solar neighborhood), the stars hosting Neptune-sized planets investigated by \citet{2012Natur.486..375B} could be more metal-rich than non-hosts. Further, the \citet{2012Natur.486..375B} sample contains no stars with \teff \,$<4500$~K, and cannot draw conclusions about the role of metallicity on the frequency of Neptune-sized planets around late K and M dwarfs.

\citet{2004ApJ...612L..73L} and \citet{Adams:2005ys} argued that the core-accretion model of planet formation predicts that late K and M dwarfs have significantly fewer giant planets than their solar-type counterparts. Disks around M dwarfs have longer dynamical (orbital) times \citep[resulting in slower planet growth,][]{1996Icar..124...62P}, lower surface densities \citep{Hartmann:1998fk, Scholz:2006lr}, and less total disk mass \citep{Williams:2011qy} than those around their solar-mass counterparts. \citet{2004ApJ...612L..73L} and \citet{Adams:2005ys} predicted that although M dwarfs should have fewer giant planets, Neptune-like objects and terrestrial-type planets should be common around such stars.

There are theoretical reasons to suspect the presence of Neptunes around M dwarfs should be correlated with stellar metallicity, even if this is not the case for FGK dwarfs. Numerical simulations indicate that the initial surface density of solids in a disk (for which stellar metallicity is a proxy) controls the mass and number of planets. \citet{Kokubo:2006lr} found that the mass of the largest and second largest planet in a planetary system should scale almost linearly with the disk surface density and that the total number of planets decreases with surface density, even in the absence of giant planets. Because the surface density of solids in a planet-forming disk should scale with the metallicity, their results suggest that metal-rich systems host larger (non-Jovian) but fewer planets. Since the disk mass scales roughly linearly with the stellar mass \citep[although with considerable scatter;][]{Williams:2011qy}, it is possible that even metal-poor FGK dwarf disks have sufficient solid material to produce Neptune sized planets, as was found observationally \citep{2008A&A...487..373S, 2012Natur.486..375B}. \citet{Schlaufman:2010qy} asserted that because of the smaller disk mass around late K and M dwarfs, metallicity is more critical for the formation of Neptunes around these stars. They claimed this requirement should manifest itself as a correlation between the presence of Neptune-sized planets and the metallicity of late K or M dwarfs. 

Indeed, \citet{Schlaufman:2011pd} found that late K and M dwarfs hosting super-Earth- to Neptune-sized transiting planet candidates from \kepler have redder $g-r$ (for a fixed $J-H$) color than those with no detected transit. Based on a comparison between two open stellar clusters with different metallicities, \citet{Schlaufman:2011pd} claimed that the $g-r$ color offset is due to a difference in metallicity of $\simeq0.2$~dex, in agreement with the theoretical case laid out in \citet{Schlaufman:2010qy}.

Complicating the issue, the clusters used by \citet{Schlaufman:2011pd} to calibrate their color-metallicity relation contain very few late K and M stars. \citet{West:2004qy} and \citet{Bochanski:2013lr} found that the metallicity dependence of $g-r$ reverses sign at late K/early M spectral types, and that cooler stars have {\it bluer} $g-r$ colors if they are more metal rich. \citet{Mann:2012} explained that the origin of the $g-r$ color difference between the KOI and non-KOI sample observed by \citet{Schlaufman:2011pd} is an artifact of giant star contamination in their non-KOI sample. But the question of whether KOI M dwarfs are more metal-rich than non-KOIs remains open. 

Compared to solar-type stars, M dwarf metallicities are difficult to determine, primarily due to the presence of complex molecular lines in their visible spectra, which result in line confusion and a lack of identifiable continuum, and do not always match with current M dwarf models \citep{Allard:2011lr}. Previous techniques to determine M dwarf metallicities using color-magnitude diagrams \citep{Johnson:2009fk, Schlaufman:2010qy} require astrometric parallaxes, which are largely unavailable for \kepler targets. Visible-light spectroscopic techniques were developed \citep[e.g., the $\zeta$ index,][]{Lepine:2007fk,Dhital:2012lr}, and are a reliable indicator of whether an M dwarf is a sub-dwarf or ultra sub-dwarf. However, $\zeta$ saturates near solar metallicity \citep{Woolf:2009qy, Mann:2013gf}, making it unreliable for measuring metallicities higher than that of the Sun. \citet{2010ApJ...720L.113R} and \citet{Terrien:2012lr} demonstrated that atomic lines in the $K$- and $H$-bands (respectively) can be used to estimate metallicities for M dwarfs. They calibrated their methods using $\sim$20 wide binaries, but their samples were restricted in both spectral type (M0-M4), and metallicity ($-0.5<$\,[Fe/H]\,$<+0.4$). Recently, \citet{Mann:2013gf} (henceforth M13) used 110 wide binaries spanning K5 to M6, and $-1.04 <$\,[Fe/H]\,$< +0.56$, and derive improved calibrations to determine metallicities using visible, $J$-, $H$-, or $K$-band spectra. 

Using the techniques of \citet{Rojas-Ayala:2012uq}, \citet{Muirhead:2012pd} estimated metallicities for late K and M planet candidate hosts. \citet{Muirhead:2012pd} found that late K and M KOIs' metallicities are consistent with the solar neighborhood \citep[$\simeq-0.10$,][]{2008MNRAS.389..585C}, but did not measure the metallicity of the overall field for comparison. \citet{Dressing:2013fk} fitted $grizJHK$ colors from the Kepler Input Catalog \citep[KIC,][]{Batalha:2010fk, Brown:2011fj} to the stellar models of \citet{Dotter:2008fk} to determine $R_*$, $M_*$, \teff, and [Fe/H] for M dwarf \kepler targets.  However, the colors of late K and early-M dwarfs are usually not reliable indicators of metallicity \citep{Lepine:2013lr,Mann:2013gf}, and metallicities from \citet{Dressing:2013fk} show weak or no significant correlation with those from \citet{Muirhead:2012pd}.

In this paper we investigate wheather the size and multiplicity of planets around late-type dwarfs depend on the metallicity of the host star. In Section~\ref{sec:sample} we describe our sample of planet candidate hosts and our comparison sample of dwarfs with no detected transit. In Section~\ref{sec:obs} we detail our visible and near-infrared observations of \kepler stars. We derive a new calibration in Section~\ref{sec:metal} to determine [Fe/H] from visible wavelength spectra. We then apply this calibration, and others from M13, to calculate the [Fe/H] of the KOI and non-KOI samples. In Section~\ref{sec:results} we report the metallicity distributions of late K and M dwarf hosts of Earth-, Neptune-, and Jovian-sized planets, hosts of multiple detected planets, and dwarfs with no detected transits. In Section~\ref{sec:discussion} we conclude with a brief discussion of possible complications and the consequences of our findings.

\section{Sample}\label{sec:sample}
\subsection{KOI Sample}
We selected KOIs from \citet{Batalha:2013lr} with $K_P-J>1.85$, where $K_P$ is the magnitude in the \kepler bandpass, and $J$ is from the Two Micron All Sky Survey \citep[2MASS,][]{Skrutskie:2006lr}. This sample includes all dwarfs with $T_{eff}<4100$~K and some as warm as $4500$~K \citep{Mann:2012}. We excluded 4 KOIs that are probably false positives. KOI-977 is a a giant star \citep{Muirhead:2012pd}. KOI-1902 has a V-like transit shape, and flux variations indicative of an eclipsing binary. In the latest planet candidate release, KOI-1164 has been added to the false positive list\footnote{http://exoplanetarchive.ipac.caltech.edu/}. The light curve of KOI-256 shows no limb-darkening, which is indicative of a stellar eclipse, rather than a planetary transit. As a test, we obtained two spectra of KOI-256 6h apart (see Section~\ref{sec:obs} for a description of observations). The spectra show a radial velocity difference of $>100$\ km\,s$^{-1}$, suggesting that the transit is an eclipsing white dwarf-M dwarf binary, later confirmed by \citet{Muirhead:2013lr}. The remaining sample of KOIs contains 157 planet candidates orbiting 106 dwarfs.

\subsection{{\it Kepler} non-KOI sample}
\citet{Mann:2012} show that $>90\%$ of the bright ($K_P<14$), and $\simeq7\%$ of the faint ($K_P>14)$, late K and M ($K_P-J>2.0$) {\it Kepler} targets are giant stars. $JHK$ colors are sometimes used to identify giant stars \citep[e.g.,][]{Lepine:2011vn}, however, these colors are known to be metal-sensitive \citep{Leggett:1992lr, Muirhead:2012zr}, and the giant and dwarf branches overlap in $JHK$ color space for K type stars \citep{Bessell:1988qy}. Instead, we screen out giants stars using their reduced proper motion, defined as:
\begin{equation}
H_J = J + 5\log \mu + 5, 
\end{equation}
where $\mu$ is the total proper motion in arcseconds yr$^{-1}$ and the $J$ magnitude is taken from 2MASS. 

We computed proper motions for each \kepler target star using all available astrometry from the USNO-B1.0 \citep{Monet:2003fj}, 2MASS \citep{Skrutskie:2006lr}, and SDSS \citep{Ahn:2012kx} catalogs, using the algorithm described in \citet{Kraus:2007yq}. We obtained the astrometry for each star from the Vizier archive using the IDL routine queryvizier.pro \citep{Landsman:1993kx}, and then combined the astrometry epochs from all surveys using a weighted least-squares fit. Our algorithm tested the goodness of each fit for each proper motion and rejected all astrometry outliers at $> 3 \sigma$. Most of these outliers were found in the photographic survey data, not in 2MASS or SDSS, due to the heavy weight assigned to the modern CCD-based epochs. The resulting catalog has proper motion uncertainties a factor of $\sim$2 smaller than those from USNO-B alone ($\sim$3 mas~yr$^{-1}$ versus 6--7 mas~yr$^{-1}$). 

For bright ($R \la 12$) stars, many of which are saturated in one or more of the above surveys, we adopted proper motions from the Third USNO CCD Astrograph Catalog \citep[UCAC3,][]{Zacharias:2010lr}. UCAC3 extends to $R = 16$, though the proper motion errors become quite large at $R > 13$--14. The typical errors in the UCAC3 proper motions are $\sim$1--3 mas~yr$^{-1}$ for stars as faint as $R\simeq12$ and $\sim$6 mas~yr$^{-1}$ for those as faint as $R \simeq16$. 

\begin{figure}[htbp] 
   \centering
   \includegraphics[width=0.5\textwidth]{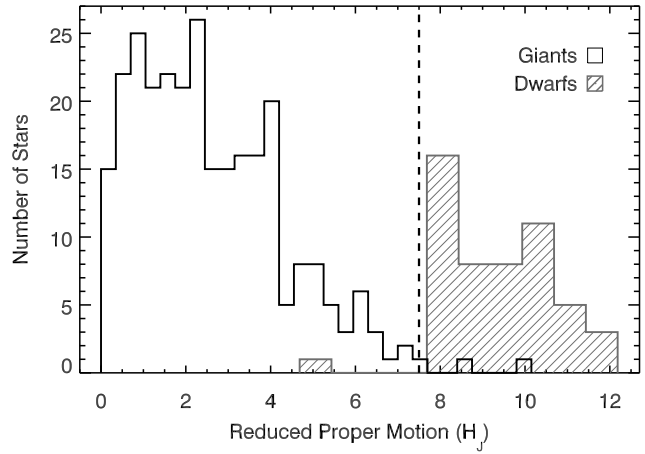} 
   \caption{Distribution of reduced proper motions for spectroscopically confirmed late K and M giants (black) and dwarfs (grey, hashed) from \citet{Mann:2012}. Stars with questionable proper motions (error in proper motion $>25$ mas yr$^{-1}$) were excluded. We utilized a reduced proper motion cut of $H_J>7.5$ (shown as a vertical dashed line) to remove interloping giant stars from our non-KOI sample.}
   \label{fig:propmotion}
\end{figure}

We use stars with known luminosity class from \citet{Mann:2012} to test possible reduced proper motion cuts. Figure~\ref{fig:propmotion} shows the reduced proper motions for the giant and dwarf samples, excluding those stars for which the errors in total proper motion are $>25$ mas yr$^{-1}$ and those that had potential contamination from a nearby star. A proper motion cut of $H_J>7.5$ excludes only 1 dwarf (of 52) in the sample, and includes only 3 giant stars (of 278). 

To establish the metallicity of \kepler target late K and M dwarfs, we randomly selected 100 stars observed by \kepler in Quarters 1 through 8 that have no detected planets (non-KOIs) for NIR spectroscopy with the criteria $K_P-J>1.85$ and $H_J>7.5$ (see Section~\ref{sec:obs} for a description of the observations). Spectra of three of these 100 contain strong CO features at $\sim$2.35~\um\ indicative of giant stars \citep{Cushing:2005lr,Rayner:2009kx}, and were excluded from our analysis. 

\section{Observations \& Reduction}\label{sec:obs}
Spectra of KOIs were obtained with the SuperNova Integral Field Spectrograph \citep[SNIFS,][]{Lantz:2004} on the University of Hawaii 2.2m telescope atop Mauna Kea. SNIFS covers 3200\,\AA\ to 9700\,\AA\ at a resolution of $1000 < R < 1300$.  SNR was $>80$ redward of 6000\,\AA\ for each target. The SNIFS processing pipeline \citep{Bacon:2001} automatically performed basic data reduction. This included dark, bias, and flat-field corrections, removing of bad pixels and cosmic rays, and sky subtraction. The SNIFS pipeline used arcs taken at the same position as the target to wavelength-calibrate the data. Spectrophotometric standards from \citet{Oke:1990}, taken over the course of each night, were used in conjunction with a model of the atmosphere above Mauna Kea \citep{Buton:2013lr} to correct for instrument response and atmospheric extinction. We shifted the spectra in wavelength to the rest frames of their emitting star by cross correlating them with a similar spectral type template from the Sloan Digital Sky Survey \citep{Bochanski:2007lr}.

We obtained near-IR spectra of the 100 non-KOI targets using the SpeX spectrograph \citep{Rayner:2003lr} attached to the NASA Infrared Telescope Facility (IRTF) on Mauna Kea. SpeX observations were taken in the short cross-dispersed mode using the 0.3$\arcsec$ slit, which yielded a resolution of $R\simeq2000$ from 0.8 to 2.4\um. SNR in the $H$- and $K$-bands was  typically $>80$. To correct for telluric lines, we observed an A0V-type star within 30 minutes of time and 0.1 airmass of the target observation. To remove effects from large telescope slews, we obtained flat-field and argon lamp calibration sequences after each A0V star. Spectra were extracted and reduced using the SpeXTool package, which performed flat-field correction, wavelength calibration, and sky subtraction \citep{Cushing:2004fk}. Telluric corrections were computed from each A0V star using the {\it xtellcor} package \citep{Vacca:2003qy}, and then applied to the relevant target spectra. We then placed each spectrum in the star's rest frame by cross-correlating it to a spectrum of a template star (of a similar spectral type) from the IRTF spectral library \citep{Cushing:2005lr, Rayner:2009kx}. 

\section{Determination of [Fe/H] and $R_*$}\label{sec:metal}
M13 provide empirical calibrations for calculating [Fe/H] from indices in visible, $J$-, $H$-, or $K$- band spectra of late K and M dwarfs. However, M13 have higher SNR observations (SNR\ $\simeq150$) than those obtained for the much fainter targets observed in this program. As a result, when we applied the calibrations from M13 on features blueward of 6000\,\AA\ (NIR calibrations are less affected) the resulting errors were large ($\gtrsim0.1$~dex) from measurement noise alone.

\begin{figure}[htbp] 
   \centering
   \includegraphics[width=0.45\textwidth]{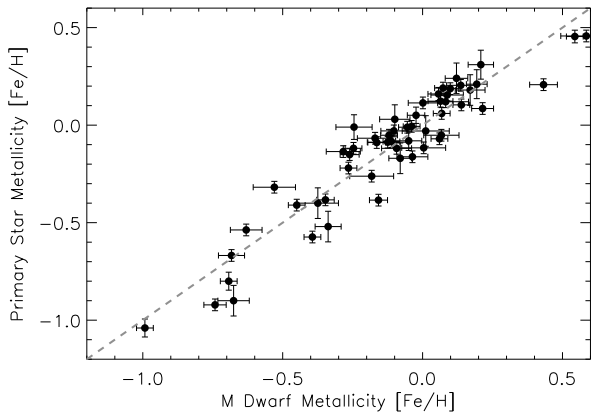} 
   \caption{Metallicity of the primary (FGK) dwarf from M13 versus metallicity for the late K or M dwarf companion derived using Eqn.~\ref{eqn:feh}. The dashed line indicates equality. The binary sample covers the full range of metallicities and spectral types in our KOI and non-KOI sample.}
   \label{fig:newcal}
\end{figure}

To mitigate SNR errors, we took the calibrator (wide binary) sample of M13 and repeated their process of defining a visible wavelength metallicity calibration. However, we restricted ourselves to indices redward of 6000\,\AA\ where the SNR of our observations is the highest. We then derived the following calibration:
 \begin{eqnarray}\label{eqn:feh} \nonumber 
\mathrm{[Fe/H]} &=& 0.68F1 + 0.53F2 - 0.32F3 \\ && - 1.0\mathrm{Color1} - 0.26,  
\end{eqnarray}
where Color1 is a temperature sensitive index from \citet{Hawley:2002fk}, and F1, F2, and F3 correspond to the equivalent widths of features at 8191-8225\,\AA, 8860-8880\,\AA, and 9179-9199\,\AA, respectively\footnote{Note: all wavelengths in this work are reported in vacuum}. We used the pseudo-continuum regions defined in M13. Figure~\ref{fig:newcal} shows the primary star metallicity as a function of the K/M dwarf metallicity derived from Eqn.~\ref{eqn:feh}. Equation~\ref{eqn:feh} has a root mean square error (RMSE) of 0.08~dex, and an adjusted square of the multiple correlation coefficient ($R_{\mathrm{ap}}^2$) of 0.82, indicating it is roughly as reliable as the corresponding calibration from M13 (RMSE=0.07 and $R_{\mathrm{ap}}^2$=0.84).

Metallicities of the non-KOI sample were calculated as the weighted means of the $J$-, $H$-, and $K$-band calibrations outlined in M13. Weights were based on the measurement errors in each band added in quadrature with the errors from each calibration (e.g., the $J$-band calibration has a significantly higher RMSE than the $H$- and $K$-band calibrations). 

Using metallicities from two different sources (visible and NIR) engenders the risk of systematic differences. However, the empirical relations we utilized are calibrated using an identical set of wide binaries. As a check, we selected a sample of 55 late K to mid-M dwarfs from \citet{Lepine:2013lr} that have both visible wavelength spectra from SNIFS and NIR spectra from SpeX. The scatter between [Fe/H] values derived using visible and NIR spectra are consistent with combined measurement and calibration errors for these two techniques ($\sigma=0.14$~dex). More importantly, there is no significant offset between metallicities derived from the two methods (median difference in metallicity is $0.01\pm0.02$~dex). 

\begin{figure}[htbp] 
   \centering
   \includegraphics[width=0.45\textwidth]{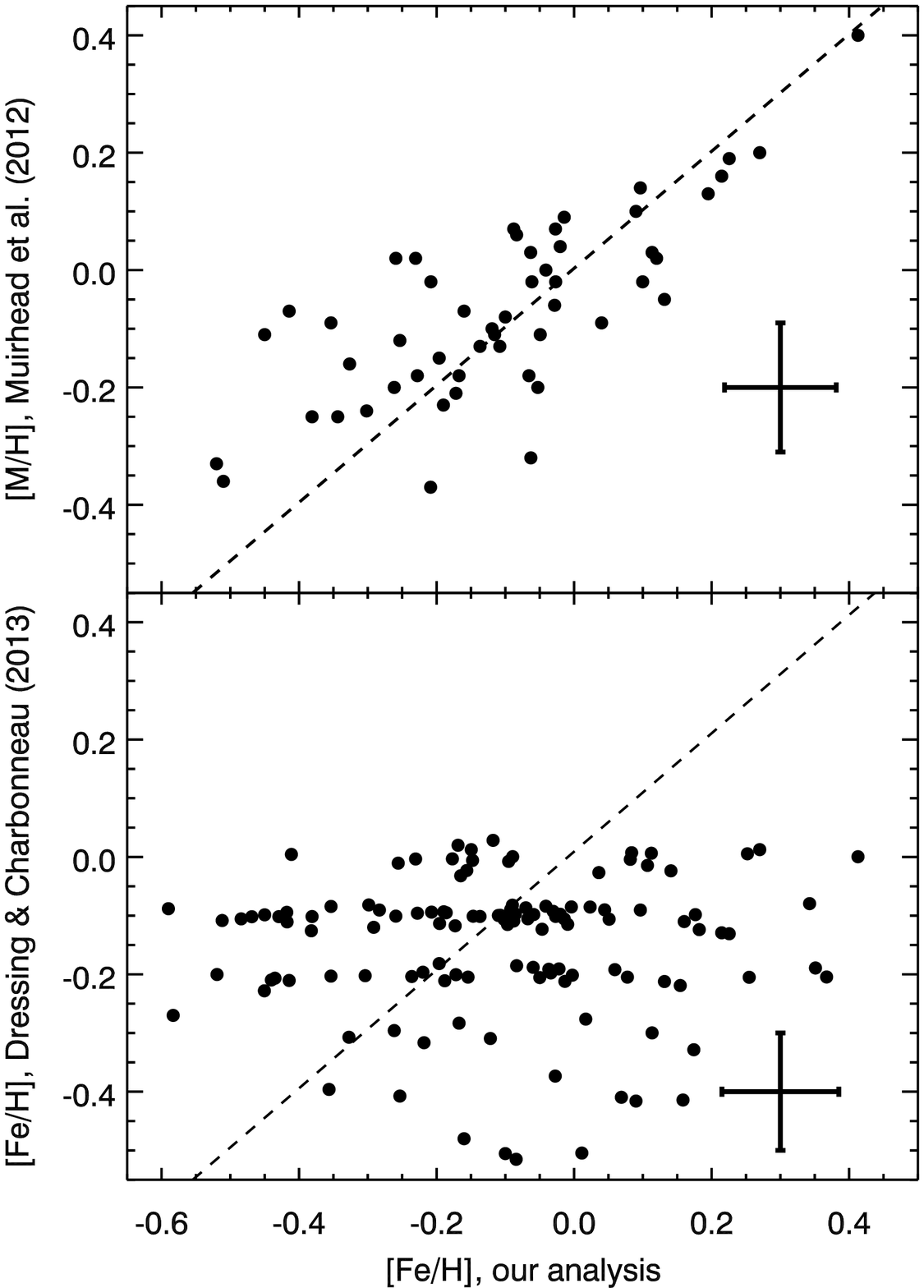} 
   \caption{Metallicities from \citet{Muirhead:2012pd} (top) or from \citet{Dressing:2013fk} (bottom) as a function of those derived from our own program. The dashed lines indicate equality. We have added artificial scatter ($\simeq0.01$~dex) to metallicities from \citet{Dressing:2013fk} for clarity. Typical errors are shown in the bottom right of each plot. Note that \citet{Muirhead:2012pd} use [M/H] instead of [Fe/H]. Our metallicities are mostly within 1$\sigma$ (reduced $\chi^2\simeq1$) of those from \citet{Muirhead:2012pd}, with the exception of those with [Fe/H]\,$<-0.3$ which are more discrepant. Our [Fe/H] values greatly differ (reduced $\chi^2>3$) from those of \citet{Dressing:2013fk}.}
   \label{fig:comparison}
\end{figure}

We compare our metallicities of our KOI sample with values from \citet{Muirhead:2012pd} in Fig.~\ref{fig:comparison}a. Our metallicities are consistent with their's for stars with [Fe/H]\,$>-0.3$. However, for more metal-poor stars, metallicities from our analysis are systematically lower and the scatter between metallicity estimates is higher. Note that we report [Fe/H] values while \citet{Muirhead:2012pd} uses [M/H]. Increasing [$\alpha$/Fe] with decreasing [Fe/H] may, in part, explain this discrepancy. The calibrator (wide binary) sample of \citet{Rojas-Ayala:2012uq}, on which \citet{Muirhead:2012pd} is based, has a paucity of stars with [Fe/H]\,$<-0.5$. The result is that their calibration assigns metallicities that are systematically too high for stars with [Fe/H]\,$<-0.4$. As can be seen in Fig.~\ref{fig:newcal}, our calibration performs well for dwarfs even with [Fe/H]\,$<-0.5$.

\begin{deluxetable}{l r l l l l l l}
\tablecaption{KOI Stellar Parameters}
\tablewidth{0pt}
\tablehead{
\colhead{KOI} & \colhead{\kepler ID} & \colhead{[Fe/H]} & \colhead{$\sigma_{\mathrm{[Fe/H]}}$} & \colhead{$R_*$} & \colhead{$\sigma_{R_*}$}\\
\colhead{} & \colhead{} & \colhead{} & \colhead{} & \colhead{$R_\odot$} & \colhead{$R_\odot$}
}
\startdata
227 &    6185476 & $-0.21$ &  0.08 & 0.60 & 0.03\\
247 &   11852982 & $+0.11$ &  0.08 & 0.53 & 0.04\\
248$^a$ &    5364071 & $+0.13$ &  0.05 & 0.57 & 0.04\\
249 &    9390653 & $-0.30$ &  0.08 & 0.37 & 0.06\\
250$^a$ &    9757613 & $-0.10$ &  0.08 & 0.59 & 0.03\\
251$^a$ &   10489206 & $-0.03$ &  0.08 & 0.48 & 0.05\\
252 &   11187837 & $-0.08$ &  0.09 & 0.50 & 0.05\\
253 &   11752906 & $+0.41$ &  0.07 & 0.55 & 0.04\\
254 &    5794240 & $+0.27$ &  0.08 & 0.49 & 0.05\\
255 &    7021681 & $-0.23$ &  0.09 & 0.55 & 0.04\\
314$^a$ &    7603200 & $-0.07$ &  0.08 & 0.52 & 0.04\\
430 &   10717241 & $-0.12$ &  0.09 & 0.64 & 0.03\\
463 &    8845205 & $-0.35$ &  0.08 & 0.36 & 0.06\\
478 &   10990886 & $+0.20$ &  0.08 & 0.48 & 0.05\\
503 &    5340644 & $-0.11$ &  0.09 & 0.62 & 0.03\\
\enddata
\label{tab:kois}
\tablecomments{Table \ref{tab:kois} is published in its entirety in the electronic edition of the {\it Astrophysical Journal}, and can be downloaded with the arXiv version of the manuscript. A portion is shown here for guidance regarding its form and content.}
\tablenotetext{a}{Multi-planet candidate system as listed in the \citet{Batalha:2013lr} catalog.}
\end{deluxetable}

\begin{deluxetable}{l l l l l l l}
\tablecaption{non-KOI Stellar Parameters}
\tablewidth{0pt}
\tablehead{
\colhead{\kepler ID} & \colhead{[Fe/H]} & \colhead{$\sigma_{\mathrm{[Fe/H]}}$} \\
}
\startdata
1721911 &  $+0.37$ &  0.07\\
1996399 &  $+0.07$ &  0.06\\
2010738 &  $+0.21$ &  0.12\\
2850521 &  $-0.68$ &  0.11\\
3233853 &  $-0.01$ &  0.05\\
3342894 &  $+0.17$ &  0.09\\
3533220 &  $-0.19$ &  0.06\\
3935942 &  $+0.04$ &  0.05\\
4543236 &  $-0.19$ &  0.06\\
4543619 &  $-0.55$ &  0.12\\
4553205 &  $+0.08$ &  0.05\\
4682420 &  $-0.39$ &  0.12\\
5000970 &  $-0.26$ &  0.07\\
5165017 &  $-0.02$ &  0.06\\
5252367 &  $-0.18$ &  0.13\\
\enddata
\label{tab:kics}
\tablecomments{Table \ref{tab:kics} is published in its entirety in the electronic edition of the {\it Astrophysical Journal}, and can be downloaded with the arXiv version of the manuscript. A portion is shown here for guidance regarding its form and content.}
\end{deluxetable}

In Fig.~\ref{fig:comparison}b we compare metallicities from \citet{Dressing:2013fk} with ours for all overlapping (KOI and non-KOI) targets. \citet{Dressing:2013fk} [Fe/H] values are inconsistent with (reduced $\chi^2>3$), and show no correlation with our values. \citet{Dressing:2013fk} themselves note significant disagreement with their metallicities and those reported in \citet{Muirhead:2012pd}, highlighting the difficulties of estimating M dwarf metallicities from photometry and stellar models alone. 

We calculated stellar radii using the \teff{}-$R_*$ relationship given in \citet{Boyajian:2012lr}. We determined \teff{} for each KOI by fitting BT-SETTL models \citep{Allard:2011lr} to our visible wavelength spectra following the technique outlined in \citet{Lepine:2013lr}, except that we only included models with metallicities $\le2\sigma$ different from those we derived from Eqn.~\ref{eqn:feh}. Stellar radii in \citet{Batalha:2013lr} are based on temperatures in the KIC \citep{Brown:2011fj} and Yonsei-Yale isochrones \citep{2004ApJS..155..667D}. However, radii from \citet{2004ApJS..155..667D} are known to be inaccurate for late K and M dwarfs \citep{Boyajian:2012lr}, and KIC temperatures have been shown to be too high for late K and M dwarfs \citep{Mann:2012}. Instead, \citet{Boyajian:2012lr} derive their \teff{}-$R_*$ using empirical measurements of radii of K and M dwarfs from long-baseline interferometry. They obtain [Fe/H] values for their targets from the literature, and find no discernible metallicity dependence in their \teff-$R_*$ relation. This contradicts stellar evolutionary models such as \citet{Dotter:2008fk}, which show a strong metallicity dependence for the \teff{}-$R_*$ relation for late K and M dwarfs. We chose to use the relation from \citet{Boyajian:2012lr}, rather than the \citet{Dotter:2008fk} models, because \citet{Boyajian:2012lr} is based on empirical measurements rather than evolutionary models.

Our stellar radii are 0.06$R_\sun$ larger than those of \citet{Muirhead:2012pd}. Their stellar radii for stars with \teff{}$<3900$ are consistent with ours at $\le 1.2\sigma$. The discrepancy is larger (typically $>1.5\sigma$) for warmer stars because their temperatures are underestimated: the H$_2$O index utilized by \citet{Muirhead:2012pd} to calculate \teff{} saturates at 3800$-$4000~K \citep{Rojas-Ayala:2012uq, Dressing:2013fk}. However, \citet{Muirhead:2012pd} stellar radii are still $<3\sigma$ consistent even at warm temperatures. Our stellar radii are on average 0.02$R_\odot$ larger than those of \citet{Dressing:2013fk} with no significant trend in \teff. This offset is smaller than the typical errors from our own measurements (median $\sigma_{R_*}\simeq0.04R_\odot$). Resulting \teff, $R_*$, and associated errors for KOIs in this program are listed in Table~\ref{tab:kois}.

We adopted the $R_{planet}/R_*$ values reported by \citet{Batalha:2013lr}. \citet{Dressing:2013fk} refit the \kepler light curves for M dwarf KOIs and find that there are significant problems with some of the transit parameters reported in \citet{Batalha:2013lr}. However, these problems are primarily in $a/R_*$ (where $a$ is the semi-major axis) and impact parameter, whereas the median $R_{planet}/R_*$ value from \citet{Dressing:2013fk} is only 3\% smaller than those of \citet{Batalha:2013lr}. Further, the differences in $R_{planet}/R_*$ between \citet{Dressing:2013fk} and \citet{Batalha:2013lr} are small compared to errors in $R_{planet}/R_*$ reported by \citet{Batalha:2013lr} ($\simeq 13\%$) and errors in stellar radii ($\simeq7\%$). Moreover \citet{Dressing:2013fk} only refit transits of M dwarfs, while our sample includes many late K dwarfs.  

\section{Results}\label{sec:results}
\begin{figure*}[htbp] 
   \centering
   \includegraphics[width=0.8\textwidth]{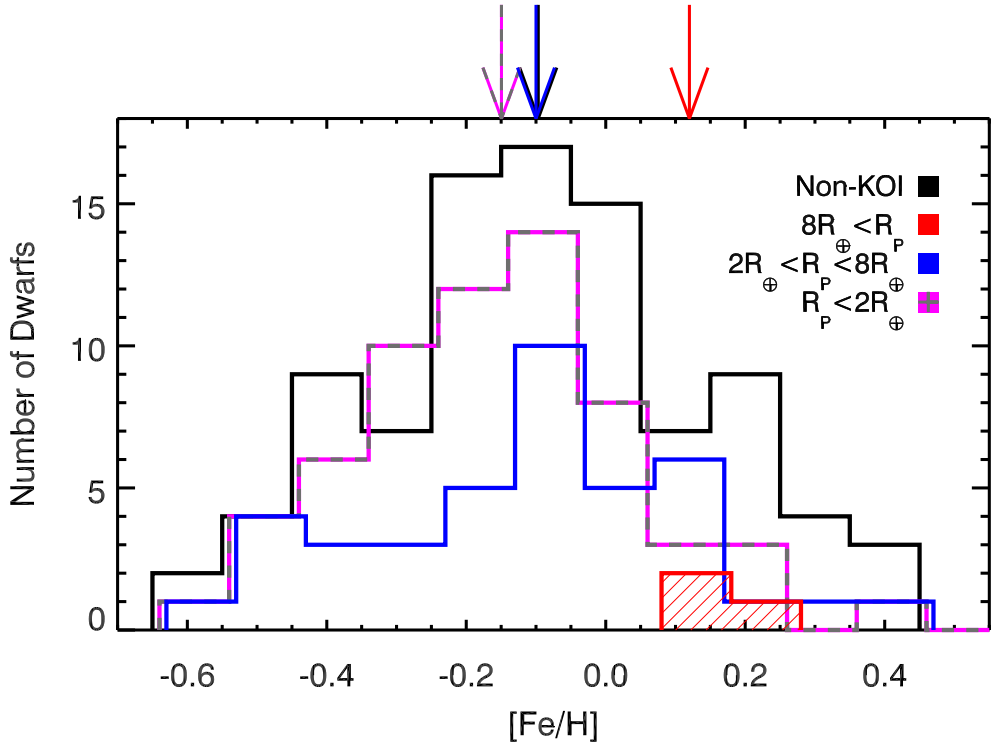} 
   \caption{Metallicity distribution of {\it Kepler} late K and M targets with no detected transit (black), Earth-sized KOIs ($R_P<2R_\earth$, magenta-grey dashed), Neptune sized KOIs ($2R_\earth<R_P<8R_\earth$, blue), and Jupiter-sized KOIs ($8R_\earth<R_P$, red, hashed). Bin sizes and locations are identical for all distributions, but the histograms are slightly offset for clarity. Arrows above the plot indicate the median of each distribution. Note that the black and blue arrows are nearly overlapping. For multi-planet systems we use the radius of the largest detected planet in the system. }
   \label{fig:metallicity}
\end{figure*}

We list [Fe/H] values and stellar radii for the KOI sample in Table~\ref{tab:kois}, and [Fe/H] values for the non-KOIs in Table~\ref{tab:kics}. Fig.~\ref{fig:metallicity} compares the metallicity distributions of dwarfs with no detected transit (non-KOI sample), as well as metallicities for different planet-size samples (Earths, Neptunes, and Jupiters). We summarize the metallicities for each distribution in Table~\ref{tab:metaldist}, and compare with the non-KOI sample using the Kolmogorov-Smirnov (KS) and Cram\'{e}r-von Mises-Anderson (CMA) tests. Systems with multiple detected planets are placed into size bins (Earth-, Neptune-, and Jupiter-size) according to the largest detected planet in the system. We also list the metallicity distribution of these multi-planet systems, according to the \citet{Batalha:2013lr} catalog, in Table~\ref{tab:metaldist}.

\begin{deluxetable}{l l l l l l l}
\tablecaption{Summary of Metallicity Distributions}
\tablewidth{0pt}
\tablehead{
\colhead{Planet-type} & \colhead{Planet sizes} & \colhead{N} & \colhead{${[\mathrm{Fe/H]}}$} & \colhead{KS$^a$} & \colhead{CMA$^b$}\\
\colhead{} & \colhead{$R_\earth$} & \colhead{} & \colhead{median} & \colhead{} & \colhead{$\sigma$}
}
\startdata
Jupiters & $R_p>8$ &  3 &  $+0.12\pm0.07$ & 2.0\% &  3.2\\
Neptunes & 2.0\,$<R_p\leq$\,8 & 40 & $-0.10\pm0.04$ & 96.9\% &  0.2\\
Neptunes2 & 2.5\,$<R_p\leq$\,8 & 17 & $-0.12\pm0.06$ & 99.5\% &  0.1\\
Neptunes3 & 3.0\,$<R_p\leq$\,8 &  6 & $-0.03\pm0.11$ & 99.5\% &  0.2\\
Earths & $R_p\leq$\,2 & 63 & $-0.15\pm0.03$ & 33.9\% &  1.2\\
Multis$^{c}$ & all & 31 & $-0.10\pm0.04$ & 39.4\% &  0.8\\
Non-KOI & not detected & 97 & $-0.10\pm0.03$ & \nodata & \nodata \\
\enddata
\label{tab:metaldist}
\tablenotetext{a}{Probability that the distribution is drawn from the same parent population as the non-KOI sample based on a Kolmogorov-Smirnov (KS) test.}
\tablenotetext{b}{Difference between distribution and non-KOI distribution (in standard deviations) as determined by the Cram\'{e}r-von Mises-Anderson statistic \citep{Anderson:1962}.}
\tablenotetext{c}{Systems with more than one transiting planet detected in the \citet{Batalha:2013lr} cataolog.}
\end{deluxetable}

Metallicities of stars hosting Jupiter-sized planets are significantly higher than those with no detected transit, which is consistent with previous findings based on radial velocity surveys of M dwarfs \citep[e.g.,][]{Johnson:2010lr}. Because there are only 3 giant planets in the sample, the difference between the Jupiter and non-KOI sample is at the edge of statistical significance, with a difference between the median [Fe/H] values of $0.22\pm0.073$ ($3.0\sigma$). 

The distribution of metallicities for dwarfs hosting Neptune-sized planets is consistent with the non-KOI sample for all metrics. We rule out the 0.2~dex offset reported by \citet{Schlaufman:2011pd} between the metallicity of stars hosting non-Jovian ($R_p\leq$\,8R$_{\earth}$) planets and the non-KOI sample at $>6\sigma$, and at $>4\sigma$ if we consider just Neptune-sized planets ($2.0\le\mathrm{R_p} \leq 8\mathrm{R_{\earth}}$).

The distribution for the Earth-sized hosts is slightly more metal poor (by 0.05~dex) than the non-KOI sample, although the offset is not significant ($\sigma=0.04$~dex). Our detected offset is consistent with predictions from \citet{Gaidos:2013qy}, who use \citet{Dotter:2008fk} models and \kepler target stars to show that M dwarfs hosting small planets have [Fe/H] values $\simeq0.02$~dex lower than those without planets, because for a given $g-r$ color, metal-poor K and M dwarfs will have smaller radii than metal-rich dwarfs (making small planets easier to detect). If the relations from \citet{Boyajian:2012lr} are correct (i.e., metallicity is negligible factor in $R_*$ for a given \teff) then this detection bias is smaller, but still present, since \citet{Boyajian:2012lr} find that the color-$R_*$ and color-\teff{} relations for K and M dwarfs have a significant metallicity dependence. 

\section{Discussion}\label{sec:discussion}
In this paper we present our comparison between metallicities of late K and M \kepler target stars and planet candidate hosts. We used techniques of (or modified techniques of) \citet{Mann:2013gf} to calculate [Fe/H]. We then investigated correlations between stellar metallicity and the presence, multiplicity, and size of any detected planets. We draw four main conclusions:
\begin{itemize}
\item The metallicity distribution of late K and M {\it Kepler} targets is indistinguishable from that of the solar neighborhood. 
\item Late K and M \kepler dwarfs hosting giant planets are more metal-rich than those without detected planets.
\item Late K and M hosts where the largest detected planet is Earth- or Neptune-sized have metallicities consistent with those dwarfs with no detected transit.
\item Late K and M dwarfs hosting multiple detected planets are not significantly more metal-rich or metal-poor than those with no detected transit. 
\end{itemize}

An important effect is the presence of non-detected planets in the control (non-KOI) sample, which dilutes any metallicity offset between the two samples. Suppose metallicity is a bimodal distribution with stars harboring planets having metallicity greater by $\Delta_{\mathrm{[Fe/H]}}$ than those without planets.  The observed metallicity offset ($O$) between the KOI sample and the non-KOI sample is:
\begin{equation}
O = \Delta_{\mathrm{[Fe/H]}}\frac{(1-f)}{(1-ft)},
\end{equation}
where, $f$ is the fraction of stars with planets, and $t$ is the probability of detecting the planet (e.g., the geometric transit probability). For transiting planets, $ft$ is small, and the denominator $\simeq1$. In the case that $f$ approaches 1, $O\simeq0$, because the non-KOI sample is completely diluted with undetected planets. This may be the case when considering Earth- to super-Earth-sized planets and all orbital periods around M dwarfs \citep{Swift:2013vn}. In the case of giant planets, which are relatively rare \citep[2\% for period $<$85 days;][]{Fressin:2013qy}, dilution is negligible and $O\simeq \Delta_{\mathrm{[Fe/H]}}$. For Neptunes, \citet{Fressin:2013qy} find that $\sim$25\% of stars harbor a Neptune ($2\,R_{\earth} <R_p\leq8\,R_{\earth}$) and periods $<$85 days (note that only 1 of the planets in our sample has a period $\gg$85~days). In this case $O\simeq 0.75\Delta_{\mathrm{[Fe/H]}}$, which has little effect on our conclusions. 

We examined how our results change as a function of how we define Earth-, Neptune-, and Jupiter-sized planets. There are no KOIs in our sample with radii between $6R_\earth$ and $9R_\earth$, so our choice of a Neptune-Jupiter boundary is unimportant. We investigated the effect of changing the Earth-Neptune boundary by considering two sub-samples; Neptunes2, defined as $2.5\,R_{\earth}<R_p\leq8\,R_{\earth}$, and Neptunes3, defined as $3.0\,R_{\earth} < R_p \leq 8\,R_{\earth}$. All three Neptune samples are consistent with each other and the non-KOI sample at 1$\sigma$, demonstrating that our results are not sensitive to how we define Earth- and Neptune-sized planets.

\citet{Fressin:2013qy} show that the \kepler planet search algorithm is missing planets that should have been detected based on their SNR. However, this only alters our results if the metallicity distribution of missing planets is significantly different than that of the detected planets. To check the effect of incompleteness, we considered a subsample with transit detections of SNR\ $>16$ as reported by \citet{Batalha:2013lr}, where \citet{Fressin:2013qy} suggest the detection efficiency of the \kepler pipeline is $\simeq100\%$. In this case, the sample shrinks from 157 planet candidates around 106 dwarfs to 132 candidates around 93 dwarfs. Most of the candidates removed by this cut are $R_p\leq$\,2$R_\earth$. We reran all analyses on this subsample and find that none of our conclusions are changed.

Our results disagree with those of \citet{Schlaufman:2011pd}, who claimed that late K and M {\it Kepler} stars hosting small planets are more metal-rich than non-hosting late K and M stars. In place of spectroscopic metallicities \citet{Schlaufman:2011pd} use $g-r$ versus $J-H$ colors, which been shown to be positively correlated with metallicity for F, G, and early-K dwarfs \citep[e.g.,][]{An:2009fk}. However, \citet{West:2004qy} and \citet{Bochanski:2013lr} see a {\it negative} correlation between $g-r$ color and metallicity for M dwarfs. The stars in our sample are in the transition region (mostly K5-M2), where the PHOENIX stellar atmosphere models predict little or no trend of $g-r$ color with metallicity \citep{Lepine:2013lr}. 

\begin{figure*}[htbp] 
   \centering
   \includegraphics[width=0.45\textwidth]{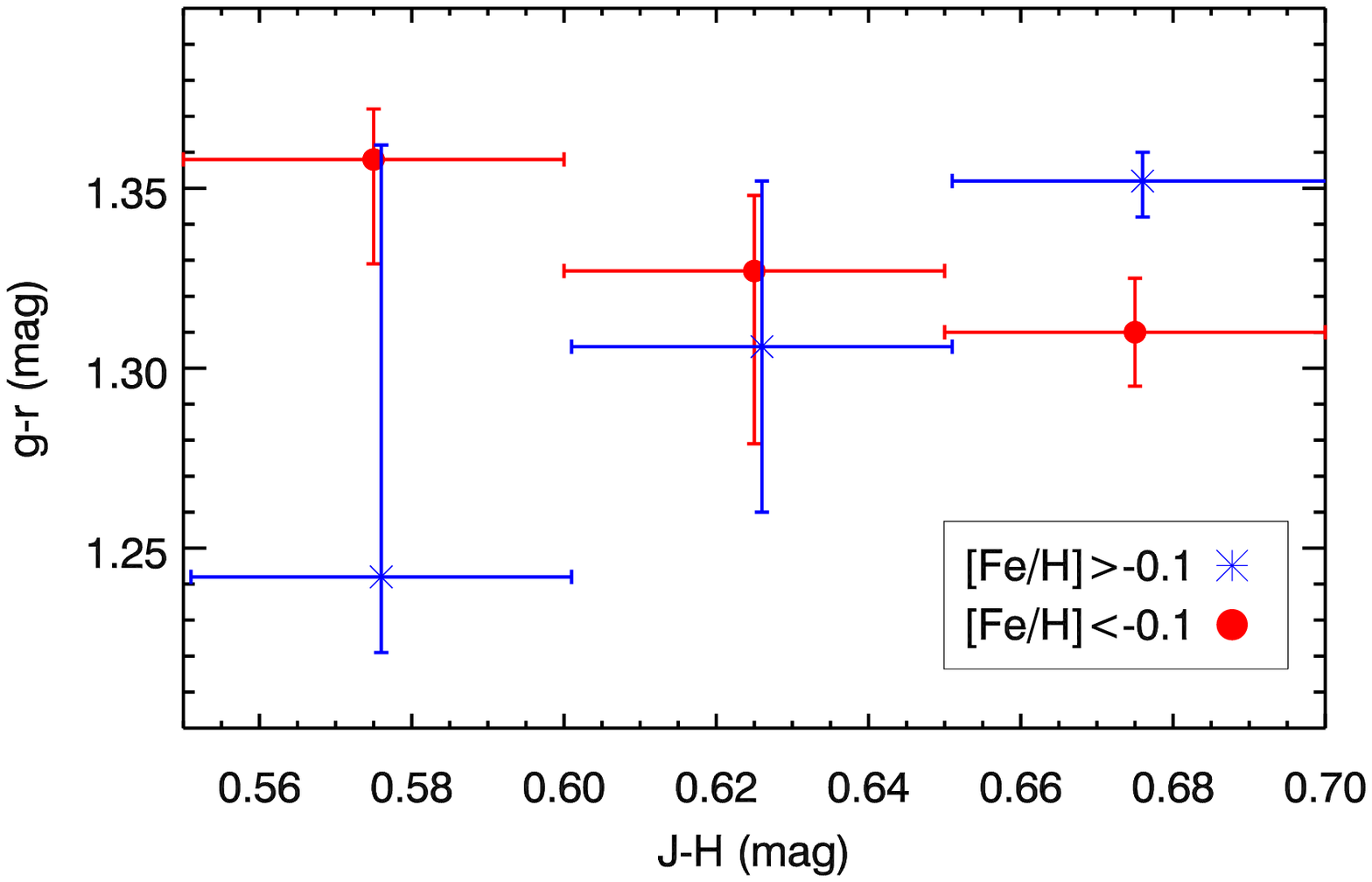} 
   \includegraphics[width=0.45\textwidth]{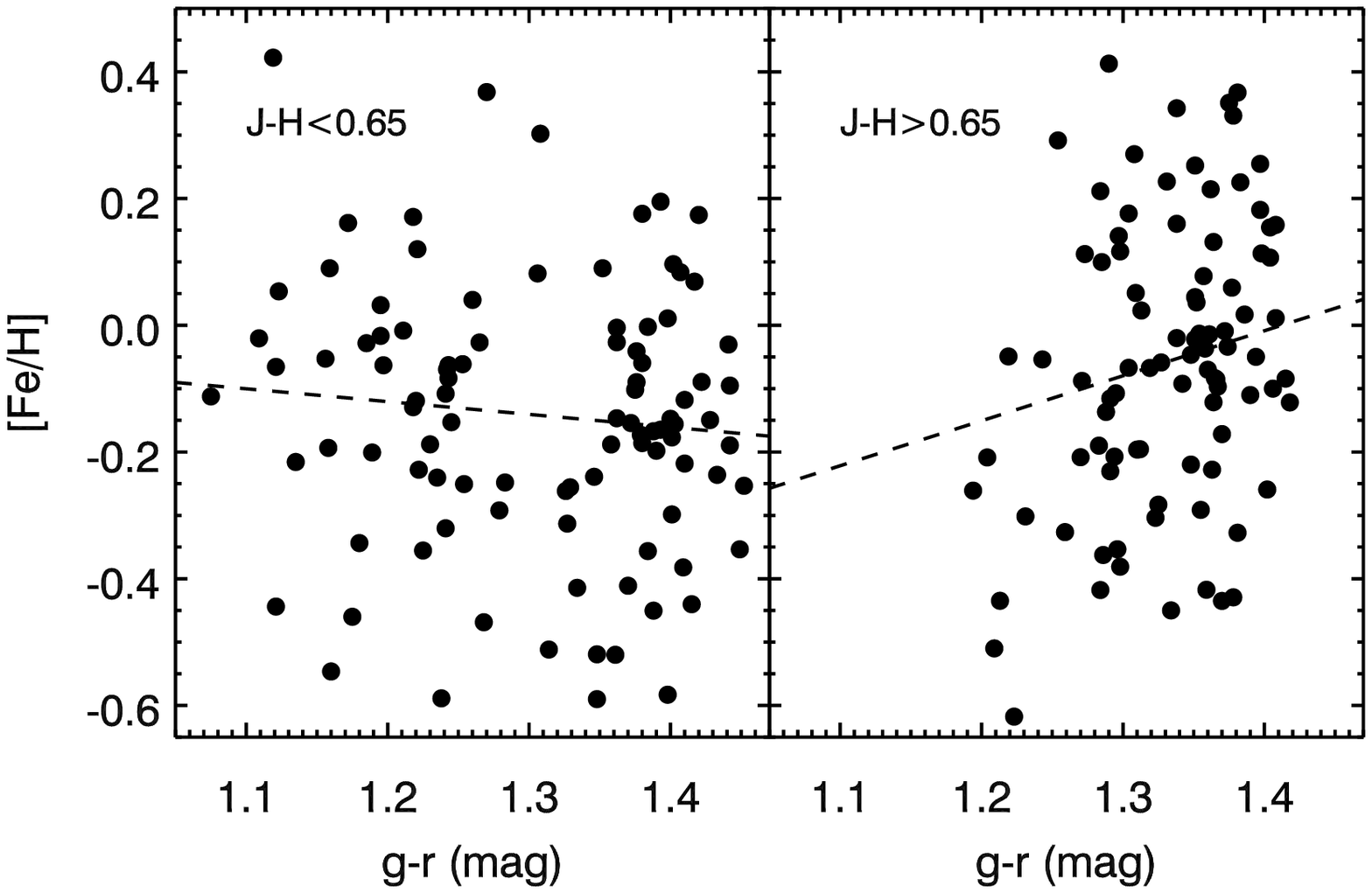} 
   \caption{Left: Median $g-r$ colors of the metal-rich (blue) and metal-poor (red) samples of \kepler dwarfs for three $J-H$ color bins centered at 0.575, 0.625, and 0.675. Bin sizes and locations are identical for all distributions, but are slightly offset for clarity. The scatter in $g-r$ colors for a given metallicity range and $J-H$ bin are determined through bootstrap resampling. For all three $J-H$ bins, the two metallicity groups are indistinguishable at 2$\sigma$. Right: [Fe/H] versus $g-r$ for two different $J-H$ bins, with the best-fit line shown. For both bins, the best fit lines are consistent with a slope of 0, and yield coefficients of determination (R$^2$) of 0.01 and 0.07, respectively. An F-test detects no significant improvement from the regression for either bin. Thus no significant correlation between $g-r$ versus $J-H$ and [Fe/H] is found. }
   \label{fig:grmetal}
\end{figure*}

We used our sample (both KOIs and non-KOIs) to investigate how $g-r$ correlates with metallicity for our range of spectral types. In Fig.~\ref{fig:grmetal}a we show the $g-r$ colors for the metal-poor ([Fe/H]\,$<-0.1$) and metal-rich ([Fe/H]$>-0.1$) samples for three different $J-H$ bins centered at $J-H$ = 0.575, 0.625, and 0.675. [Fe/H]$=-0.1$ was selected to divide the samples because this is the median metallicity of our sample. The late-type dwarf bin used by \cite{Schlaufman:2011pd} is centered at $J-H=0.62$. The distribution of $g-r$ colors of the two metallicity samples are consistent at\ $<2\sigma$ in each of the three $J-H$ bins. We use a slightly different parsing in Fig.~\ref{fig:grmetal}b, where we show [Fe/H] versus $g-r$ color for two $J-H$ bins. In each bin a least-squares linear fit to the data yield slopes that are not significantly different from 0. The fit yields coefficients of determination (R$^2$) of 0.01 and 0.07 for the $0.65>J-H$ and $0.65<J-H$ bins, respectively. An F-test comparing [Fe/H]$-\overline{\mathrm{[Fe/H]}}$ to that of [Fe/H]$-$[Fe/H]$_{fit}$ gives respective differences in the variances of only 0.50 and 0.48$\sigma$. These results strongly suggest that $g-r$ versus $J-H$ is not a good predictor of [Fe/H] for late-type dwarfs.

Our results have some important theoretical implications. Theoretical studies have suggested there is a minimum metallicity  for a protoplanetary disk to form planets \citep[e.g.,][]{Gonzalez:2001fj, Johnson:2012fr}. Assuming the metallicity of the protoplanetary disk matches that of the star later in its evolution, this suggests that low-metallicity stars should not harbor planets. \citet{Johnson:2012fr} estimate that the minimum metallicity to form a planet around a solar-type star is [Fe/H]$_{\mathrm{min}} \simeq -1.5 + \log(a)$, where $a$ is the semi-major axis in AU. For the planets in our paper ($a\lesssim0.1$) [Fe/H]$_{\mathrm{min}}$ is approximately $-2.5$. Because M dwarfs have less massive, we expect that [Fe/H]$_{\mathrm{min}}$ will be higher for these stars. Our results show that Earth and Neptune-sized planets are able to form around stars with metallicities as low as [Fe/H]$\simeq-0.6$, similar to what is seen for FGK dwarfs \citep{2012Natur.486..375B}. But it is likely that we are not probing sufficiently metal-poor dwarfs to detect the proposed planet formation threshold.

Our results also indicate that, for small planets, multiplicity is not correlated with metallicity. Interestingly, two of the four KOIs with [Fe/H]\,$<-0.5$ have $\ge3$  detected planets (KOI-961 has 3 and KOI-812 has 4), suggesting the accretion process must be efficient in collecting solids from the disk. A minimum mass solar nebula contains about 64$M_\earth$ of rock/metal/ices. Assuming that the disk mass is $\sim0.1M_*$ \citep{Williams:2011qy}, and that the amount of metals in a disk scales with its mass and the stellar metallicity, a disk around an early M dwarf with [Fe/H]\ $=-0.5$ contains $\sim10$M$_\earth$ of solids. KOI-812 (as an example) contains 4 detected planets with radii from 1.3 to 2.4R$_\earth$. Most of these are likely rocky, or are composed of rocky cores with a thin hydrogen envelopes \citep{Gaidos:2012lr}. If we assume a mass-radius relationship of $M_P\simeq R_P^2$, with $M_P$ and $R_P$ in Earth units, then the total mass in KOI-812's planets is $15.5\pm 3.1M_\earth$. Although some of this mass is hydrogen (and thus not affected by the amount of solids in the disk), this analysis does not consider undetected planets at higher semi-major axes. Thus our results suggest that the progenitors of these planets must have been very efficient in accreting most of the available solid material from the disk. 

The core accretion scenario of giant planet formation requires the formation of a $\simeq5-10\mathrm{M}_\earth$ core \citep{1996Icar..124...62P, 2005Icar..179..415H} in the $\sim2-6$~Myr timescale on which disks are observed to dissipate \citep{2001ApJ...553L.153H, 2009ApJS..181..321E}. The scarcity of giant planets around M dwarfs (especially metal-poor M dwarfs) means that either giant planets cores do not form around these stars, or that they do not form in time. The existence of objects that are likely $5-10\mathrm{M}_\earth$ (or have rocky cores of this size), even around metal-poor M dwarfs, suggests that the latter is the more viable explanation.


\acknowledgments
We thank the anonymous reviewer for helping to make the manuscript significantly better. This work was supported by NSF grant AST-0908419, NASA grants NNX10AI90G and NNX11AC33G (Origins of Solar Systems) to EG.

This paper includes data collected by the Kepler mission. Funding for the Kepler mission is provided by the NASA Science Mission Directorate. SNIFS on the UH 2.2-m telescope is part of the Nearby Supernova Factory project, a scientific collaboration among the Centre de Recherche Astronomique de Lyon, Institut de Physique NuclŽaire de Lyon, Laboratoire de Physique NuclŽaire et des Hautes Energies, Lawrence Berkeley National Laboratory, Yale University, University of Bonn, Max Planck Institute for Astrophysics, Tsinghua Center for Astrophysics, and the Centre de Physique des Particules de Marseille. Based on data from the Infrared Telescope Facility, which is operated by the University of Hawaii under Cooperative Agreement no. NNX-08AE38A with the National Aeronautics and Space Administration, Science Mission Directorate, Planetary Astronomy Program.

{\it Facilities:} \facility{IRTF}, \facility{UH:2.2m}, \facility{\kepler}

\bibliography{$HOME/dropbox/fullbiblio.bib}

\end{document}